\documentclass{llncs}

\usepackage[english]{babel}
\usepackage[utf8]{inputenc}
\usepackage{latexsym,amssymb,amstext,amsfonts,amsmath,graphicx,setspace,mathtools,bm}
\usepackage[colorinlistoftodos]{todonotes}
\usepackage{enumerate} 
\usepackage{paralist}
\usepackage{float}
\usepackage{array}

\usepackage{tikz}
\usetikzlibrary{fit,positioning}
\usetikzlibrary{calc}

\begin{document}

\title{An Empirical Study of Continuous Connectivity Degree Sequence Equivalents}
\titlerunning{Continuous Connectivity Degree Sequence Equivalents}  
%
\author{Daniel Moyer\and Boris A. Gutman \and
Joshua Faskowitz \and Neda Jahanshad\and \\ Paul M. Thompson}
\authorrunning{Daniel Moyer et al.} 
%
\tocauthor{Daniel Moyer, Boris Gutman, Joshua Faskowitz, Neda Jahanshad, and Paul Thompson}
\institute{Imaging Genetics Center, University of Southern California\\
\email{moyerd@usc.edu}}



\maketitle              

\begin{abstract}
In the present work we demonstrate the use of a parcellation free connectivity model based on Poisson point processes. This model produces for each subject a continuous bivariate intensity function that represents for every possible pair of points the relative rate at which we observe tracts terminating at those points. We fit this model to explore degree sequence equivalents for spatial continuum graphs, and to investigate the local differences between estimated intensity functions for two different tractography methods. This is a companion paper to \cite{ourselves}, where the model was originally defined.

\end{abstract}

\section{Introduction}

In the past decade, graph theoretic analyses have rapidly propagated through neuroimaging literature. Following advances in diffusion and functional MRI, the rise of connectomics has popularized the use of network representations of brain architecture and activity. Such analyses usually equate physical regions of the cortical surface with nodes in a graph, and use structural or function measurements as proxies for edge weights.

Increased use of the network representation of brain connectivity has been accompanied by the use of network statistics in disease-oriented neuroscience; nodal measures such as modularity, centrality, and degree sequences are all common descriptors for connectomes \cite{bullmore2009complex}, and the effect of various diseases on these descriptors has been the focus of recent study \cite{crossley2014hubs}. The popularity of these measures stems in part from their theoretical underpinnings. In particular, the distribution of nodal degrees has deep implications for the organization and topology of a random network.

For brain networks the choice of a particular delineation of physical regions (parcellation) and thus the choice of nodes is non-trivial. Multiple studies have shown that the choice of parcellation influences the summary measures including nodal degree for both structural and functional networks \cite{satterthwaite2015towards,wang2009parcellation,zalesky2010whole}. It remains unclear which of the many parcellations is optimal, or whether or not a single overall ``best'' parcellation exists \cite{de2013parcellation}, assuming a criterion for quality can be agreed upon. Furthermore, the study of local or multi-scale grained phenomena often require specific resolutions and thus specific parcellations. This presents a dilemma to the community: do we choose the best parcellation for their specific study, or do we choose a parcellation that generalizes well to other literature?

It is therefore valuable to explore alternative representations that avoid these issues. In particular, it is useful to construct representations of cortical connectivity that are independent of the choice of parcellation, yet still theoretically and computationally tractable both for estimation as well as statistical analysis, as well as retaining the ability to construct parcellation based connectivity. Furthermore, the exploration of continuous equivalents to currently popular network statistics and their similarities and/or deviations from current empirical observations is of interest.

In the current work we explore the empirical properties of such a model, described in a companion paper \cite{ourselves}. We explore a continuous representation of cortical connectivity that describes the observation of white matter tract\footnote{It is important to distinguish between white matter fibers (fascicles) and observed ``tracts.'' Here, ``tracts'' denotes the 3d-curves recovered from Diffusion Weighted Imaging via tractography algorithms.} endpoints using a random process defined on the gray matter/white matter interface (the inner cortical surface). This form of connectivity generalizes traditional connectomes to a parcellation-independent representation from which, given any particular parcellation, the discrete connectome may be recovered. We reproduce degree distribution results similar to those described by discrete representation analyses using our continuous model, and note the instances of discrepancy between the two. We further investigate differences between two tractography methods in the degree distribution (marginal intensity function).

\section{Continuous Connectivity Model}

In order to ensure that the reader may understand the results in Section \ref{sec:results}, we first review the theory and motivation behind this particular continuous connectivity model; more details can be found in \cite{ourselves}. In particular we introduce the framework itself and its general terminology, focusing on the key piece of our model, the Poisson point process and its intensity function. We then define the analogous statistic to degree for the continuous framework, which will serve as the empirical focus of the next section.


\subsection{Model Description and Theoretical Discussion}

A point process is a random process in which collections of discrete points are generated randomly on a measurable space. The Poisson process is the most basic of these, assuming that these points are generated independently (i.e. the appearance of one point does not affect the probability of observing another) and with some relative rate proportional to an intensity function $\lambda:\text{Domain} \rightarrow \mathbb{R}^+$. For any subset of the domain the distribution of the number of points in that subset is Poisson with parameter equal to $\lambda$ integrated over the subset. This means the expected number of points is exactly that integral. We use this process to model tract endpoint locations on the cortical surface. Our domain is the connectivity space of the cortex (the product of the gray/white matter interface with itself) which is the set of all pairs of possible endpoints, and each point in this space corresponds with a pair of tract endpoints.

In our context of connectomics, this framework produces two different structures that are both analogous to connectivity. The first is the usual region-to-region connectivity, which is produced by measuring the expected tract count on a subset of the domain which is itself composed of all pairs of points in two subsets of the cortex. These cortical subsets are not necessarily disjoint. The second representation of connectivity is the intensity function $\lambda$. While the first is an aggregation of the second, it is important to separate them. The intensity function has pointwise intensities which are not comparable with the second. The second representation generalizes the first over the choice of regions. Varying segmentation choices for the first representation can be compared in the context of the second.

A more formal definition of the framework is as follows: Let $\Omega$ be union of two disjoint subspaces each diffeomorphic to the 2-sphere representing the cortical surface for a particular hemisphere, and assume that tracts randomly and independently intersect with this surface at exactly two points. Further consider the space $\Omega \times \Omega$, which is the space of pairs of points on the cortical surface, in our case the space of all possible pairs of endpoints of tracts. We model connectivity as a function $\lambda:\Omega \times \Omega \rightarrow \mathbb{R}^+$ such that for any regions $E_1, E_2 \subset \Omega$ the number of observed tracts having one end point in $E_1$ and the other in $E_2$ is Poisson distributed with parameter 
\begin{equation}
\mathcal{C}(E_1,E_2) = \iint_{E_1,E_2}{\lambda}(x,y)dxdy.
\end{equation}

This is exactly a Poisson point process of tract endpoints over $\Omega \times \Omega$ with $\lambda$ is its intensity function. Note that for $E_1,E_2$ with non-trivial intersections this double counts the number of actual tracts observed in that intersection. If a tract has endpoint $(x,y) \in E_1 \cap E_2$, then clearly there exists a tract $(y,x) \in E_1 \cap E_2$.

We define $\lambda(x,y)$ for any particular $(x,y)$ as the \emph{pointwise connectivity} between $x$ and $y$. Though it is not technically required, we assume this process (characterized by $\lambda$) to be smooth everywhere and symmetric ($\lambda(x,y) = \lambda(y,x)$).. We further assume tracts are generated independently; more complex point process models relax this assumption, and a more general form of this model could be considered, in which $\lambda$ is vector valued, and the Poisson distribution may be substituted for any distribution parameterizable by integral terms. These models are unfortunately computationally intractable with current estimation methods, but are relevant to the proceeding discussion.

We define a \emph{regional connectivity} as the expected number of tracts between any two regions. In the Poisson case this is exactly $\mathcal{C}(E_1,E_2)$. For any partition (parcellation) $P = \bigcup_i{E_i}= \Omega $, this forms a connectivity matrix similar to the traditional connectomes (which provide connectivity for the $P\times P$ discrete space). From empirical data, a valid estimator is simply counting the number of tracts present between each pair of regions, as current methods dictate. Thus, continuous connectivity models of the form proposed here generalizes graphs generated by finite partitions of $\Omega$ (including overlapping paritions) for certain classes of edge weight distributions.

An important summary statistic of traditional graphs is the degree sequence, defined as the collection of sums of each edge for each node (this counts each edge weight exactly twice, though for different terms in the sequence). A similar construct exists for the continuous connectome which we call the \emph{marginal connectivity}, given by

\[M(x) = \int_\Omega \lambda(x,y)dy.\]

$M(x)$ itself is defined over $\Omega$. Using the assumption that $\lambda$ is continuous, it can be shown fairly easily that $M(x)$ is also continuous. Though not explored in this paper, another natural extension are marginal connectivities from a particular region. Choosing any region $E$, we can define them as $M_{E}(x) = \int_E\lambda(x,y)dy$, which is the total connectivity from point $x$ to the region $E$.

\subsection{Estimation and Asymptotic Estimator Distribution}

A sufficient statistic for Poisson process models is the intensity function $\lambda(x,y)$. Estimation of the function is non-trivial, and has been the subject of much study in the spatial statistics community \cite{diggle1985kernel}. We choose to use a non-parametric Kernel Density Estimation (KDE) approach due to an efficient closed (truncated harmonic) form for estimation. We first inflate each surface to a sphere and register them using a spherical registration (See section \ref{sec:pre-proc}), assuming each hemisphere is disjoint with the other. Our domain is then $\Omega \times \Omega$ the product of spheres $(S_1 \cup S_2) \times (S_1 \cup S_2)$. 

The unit normalized spherical heat kernel is a natural choice of kernel for $\mathbb{S}^2$. We use its truncated spherical harmonic representation \cite{chung2006heat}, defined as follows for any two unit vectors $p$ and $q$ on the 2-sphere:
\[K_\sigma(p,q) = \sum_h^H \frac{2h + 1}{4\pi}\exp\{-h(h+1)\sigma\}P^0_h(p\cdot q)\]
Here, $P_h^0$ is the $h^{th}$ degree associated Legendre polynomial of order $0$. Note that the non-zero order polynomials have coefficient zero due to the radial symmetry of the spherical heat kernel \cite{chung2006heat}. However, since we are estimating a function on $\Omega \times \Omega$, we use the product of two heat kernels as our KDE kernel $\kappa$. For any two points $p$ and $q$, the kernel value associated to a end point pair $(x,y)$ is $\kappa((p,q)|(x,y)) = K_\sigma(x,p)K_\sigma(y,q)$.

We then apply this kernel to our data in order to recover the intensity function. This is exactly $\hat{\lambda}(p,q) = \kappa((p,q)|D) = \sum_{(x_i,y_i) \in D}K_\sigma(x_i,p)K_\sigma(y_i,q)$ where $\hat{\lambda}$ denotes our estimation of $\lambda$, the true intensity. Since storing the exact functional form is expensive, in practice we evaluate this at each vertex of the cortical surface triangulated mesh. 

It should be noted that while the end result of this process is an array of values corresponding to a triangulated mesh of the cortical surface (which can be inflated into a sphere), this is not equivalent to a high resolution discrete connectome except in the roughest terms. Each point here represents a pointwise evaluation of the density function, and the difference between adjacent mesh vertices is bounded (since the domain is compact, we can show that the true intensity function must be absolutely continuous if it is continuous at all). The discrete connectome, no matter how small the region, represents an areal connectivity value, and in most cases caries no geometric information, much less a smoothness condition.

The product kernel complicates the asymptotic analysis of this estimator. Unfortunately there is no true closed form for the heat kernel on the sphere \cite{lafferty2005diffusion}, and approximations include terms that are not consistent with traditional kernel bandwidth analysis (which usually is univariate, and unbounded). On the other hand, it can be shown that the heat kernel using these approximate forms will be consistent with extant first and second moments \cite{hall1987kernel}.




\section{Empirical Results}
\label{sec:results}

\begin{figure}[h!]
\includegraphics[width=0.97\textwidth]{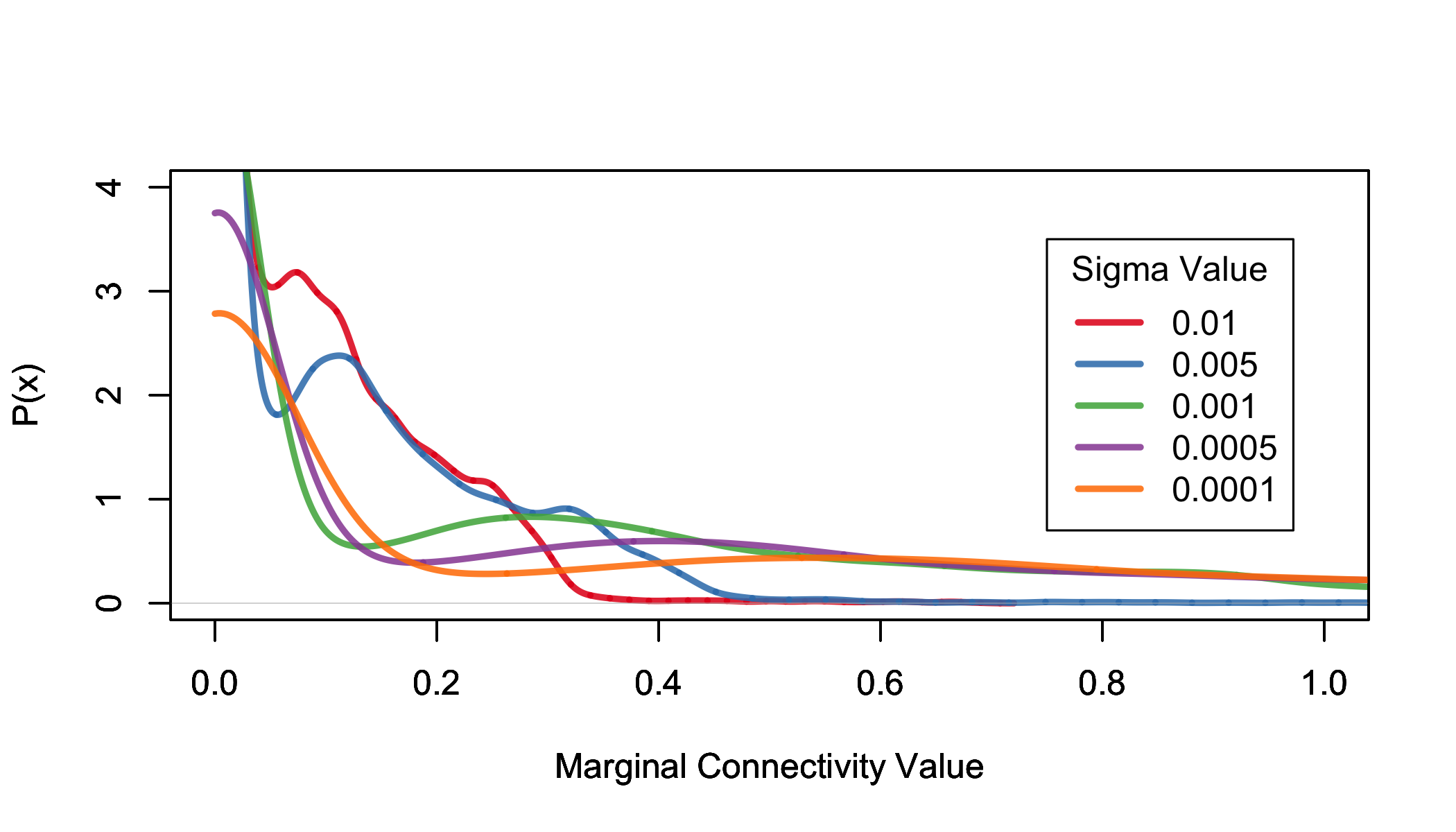}
\includegraphics[width=0.97\textwidth]{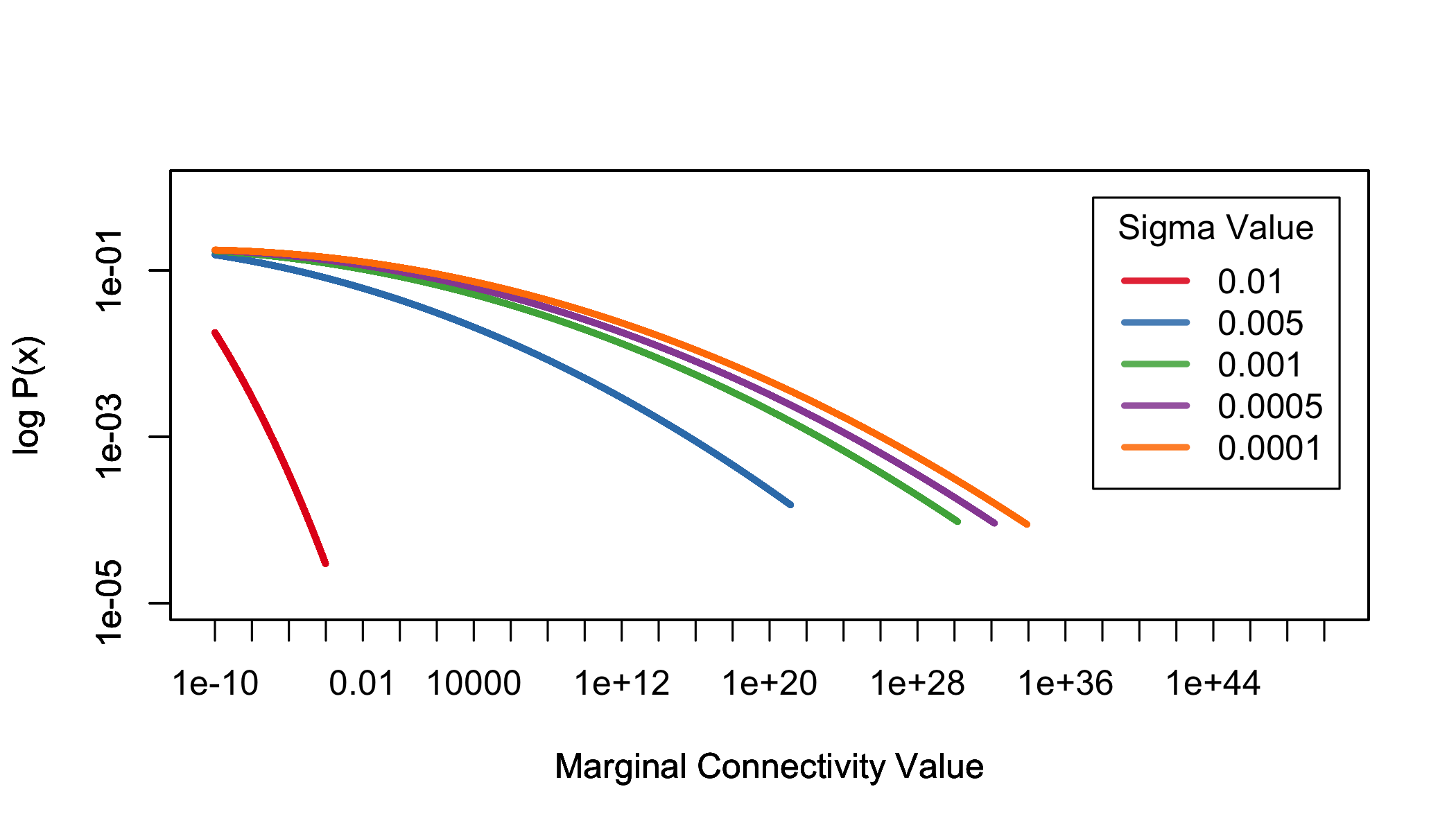}
\caption{Here we see the estimated density of sampled marginal connectivity functions. The \textbf{top} figure is the direct plot, and the \textbf{bottom} figure is the log-log plot of the same.}
\label{fig:density_ave}
\end{figure}

\begin{figure}[h]
\includegraphics[width=0.24\textwidth]{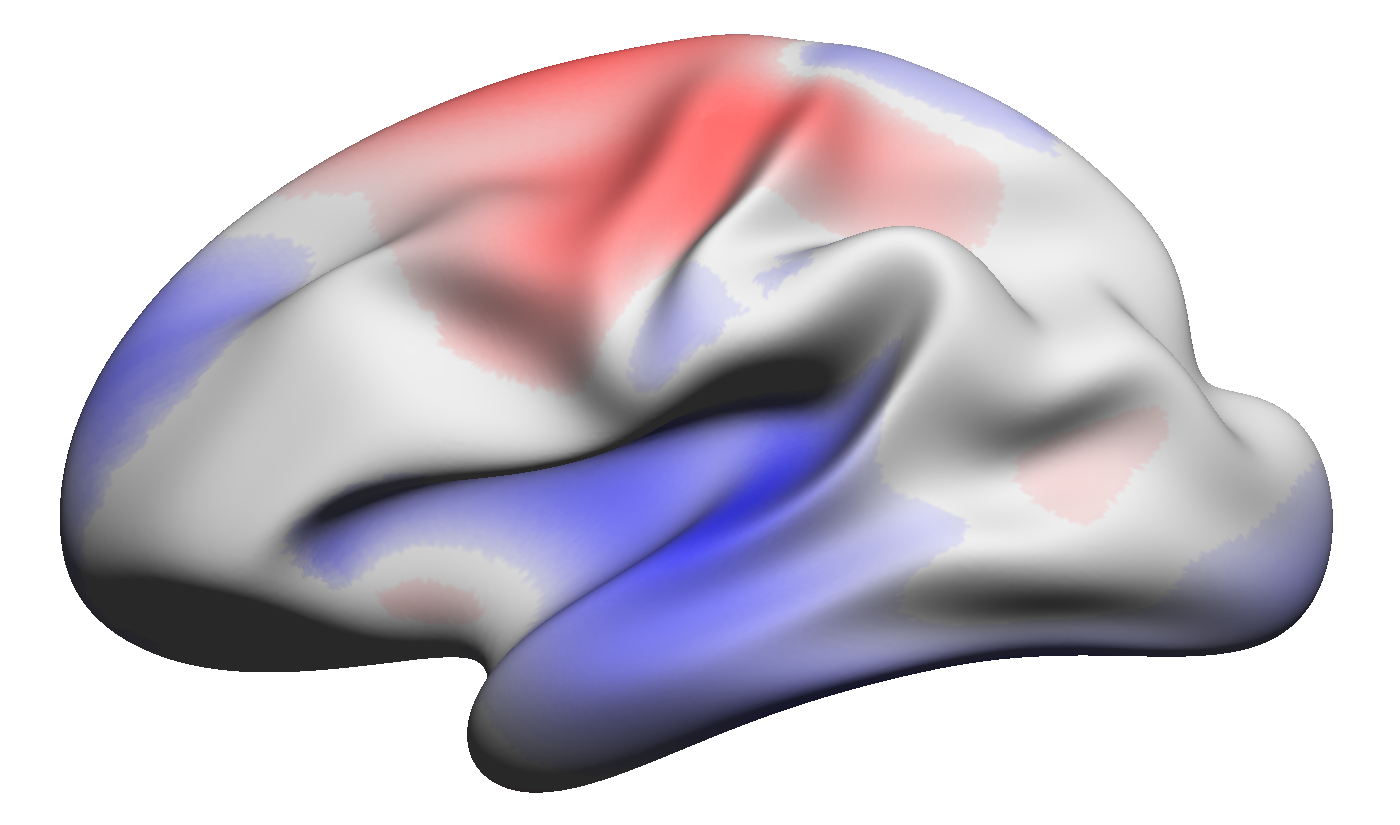}
\includegraphics[width=0.24\textwidth]{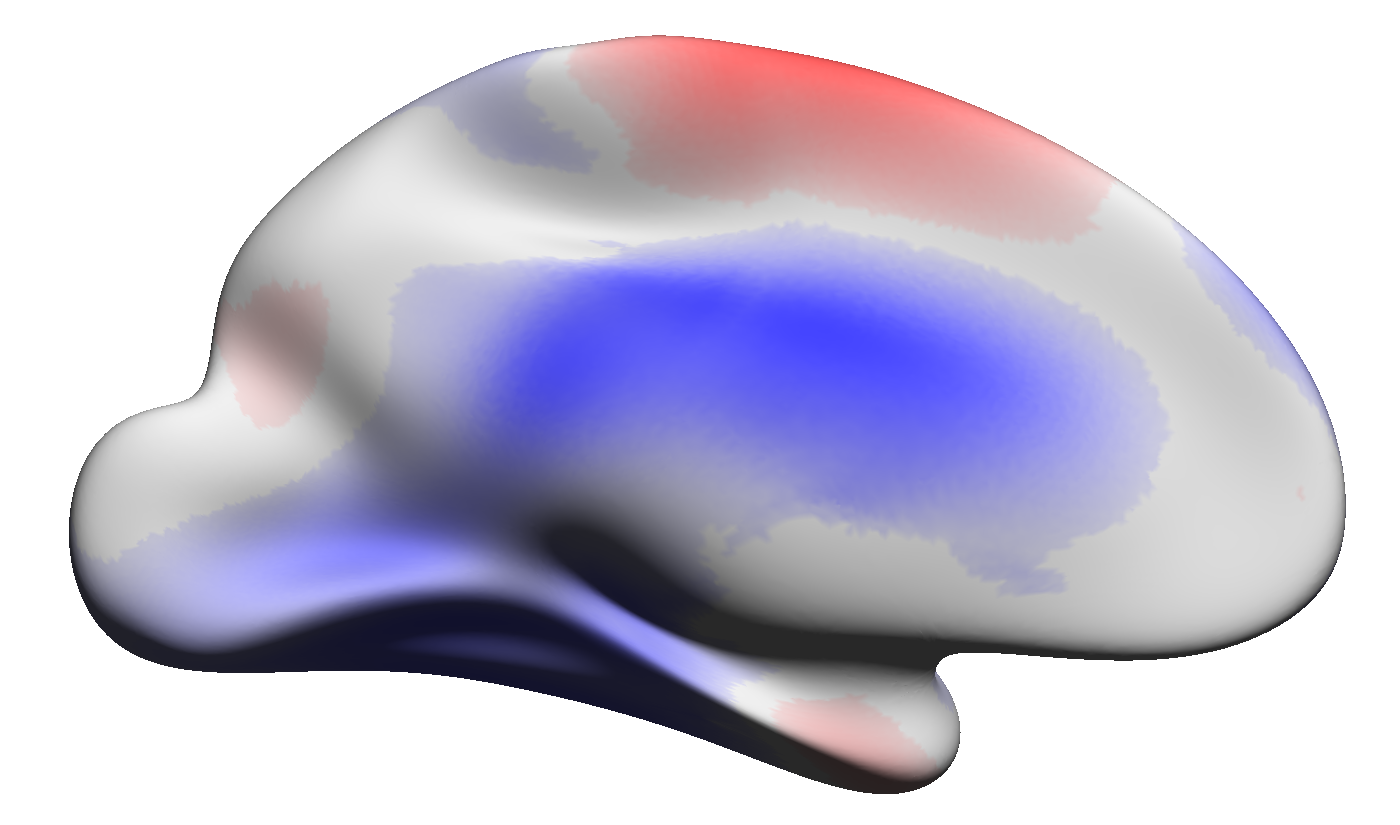}
\includegraphics[width=0.24\textwidth]{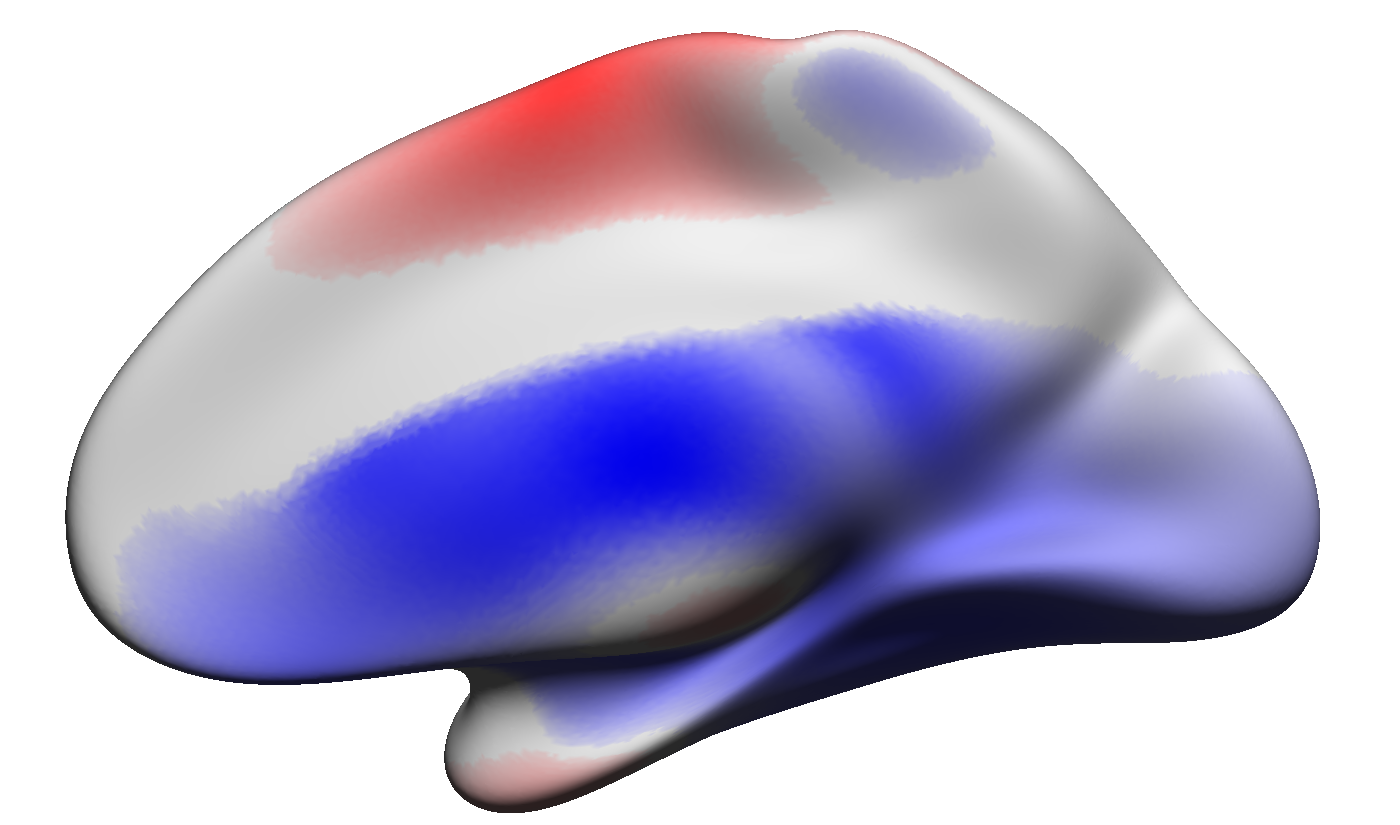}
\includegraphics[width=0.24\textwidth]{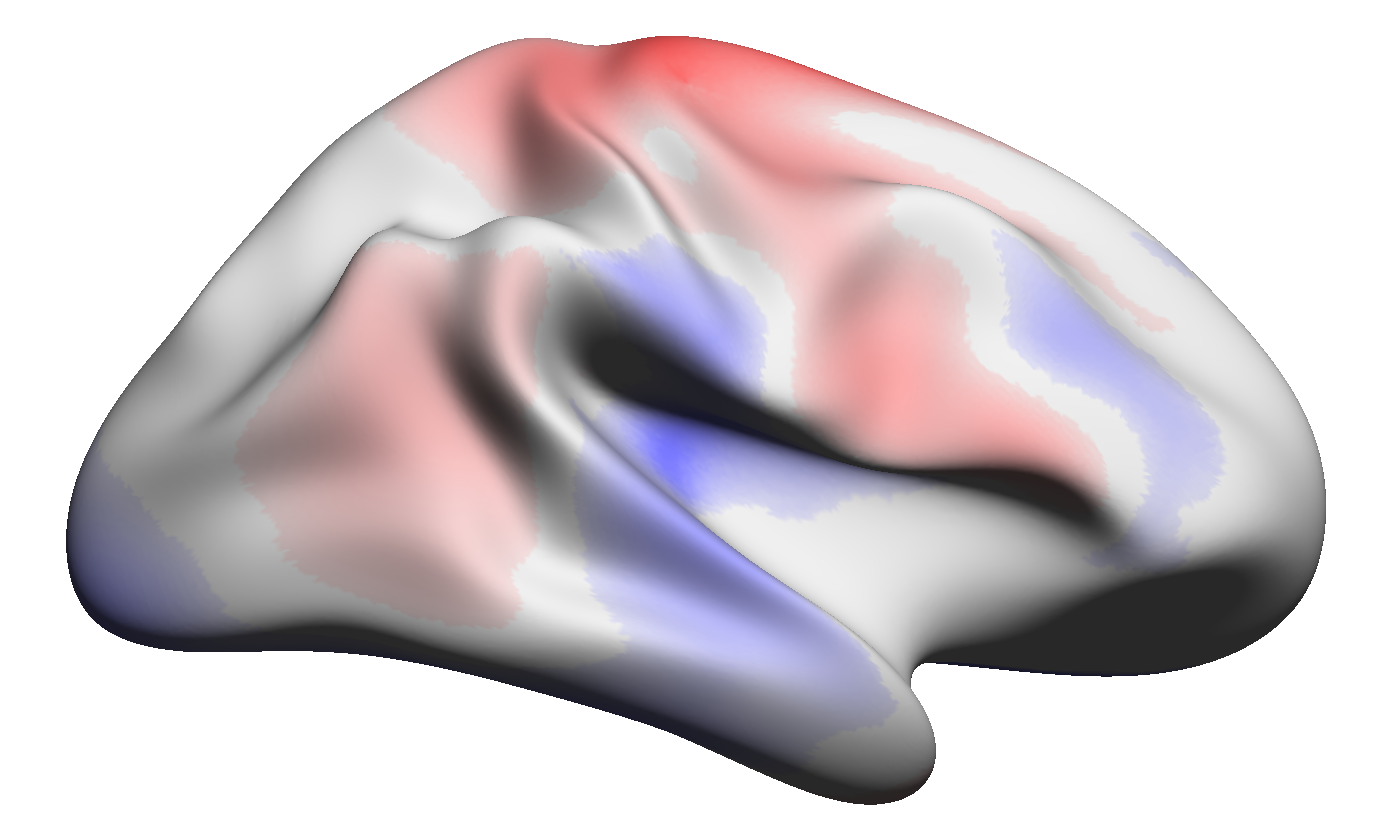}
\includegraphics[width=0.24\textwidth]{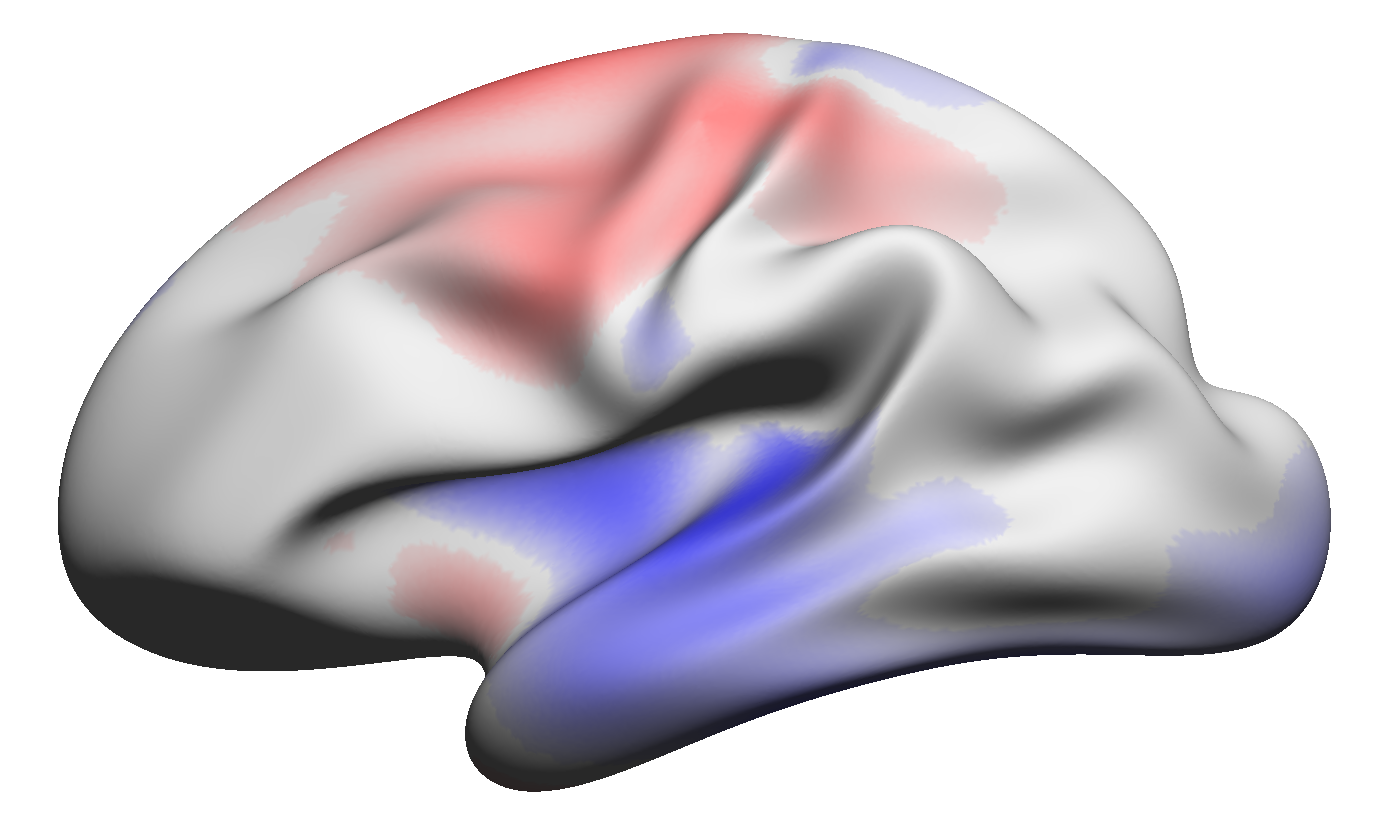}
\includegraphics[width=0.24\textwidth]{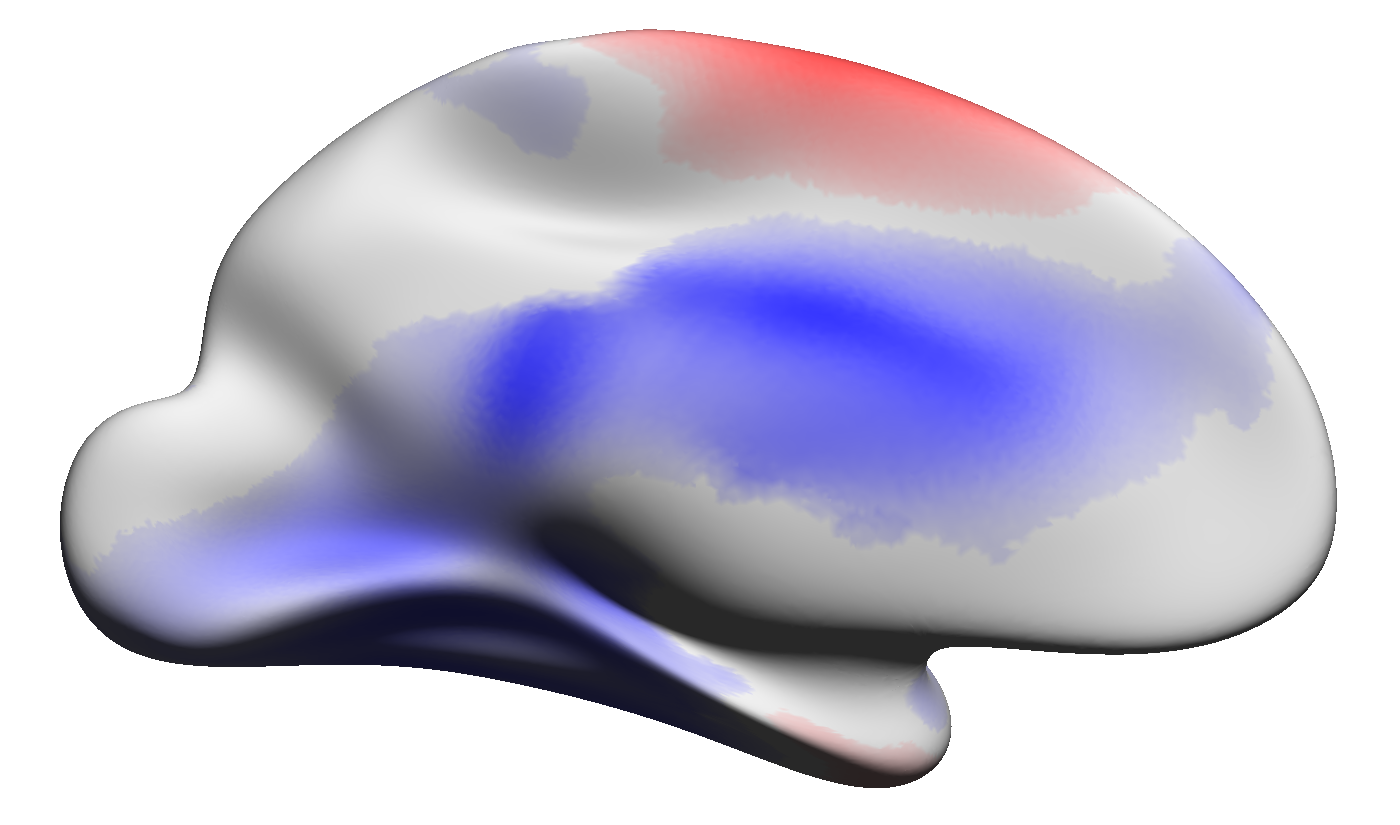}
\includegraphics[width=0.24\textwidth]{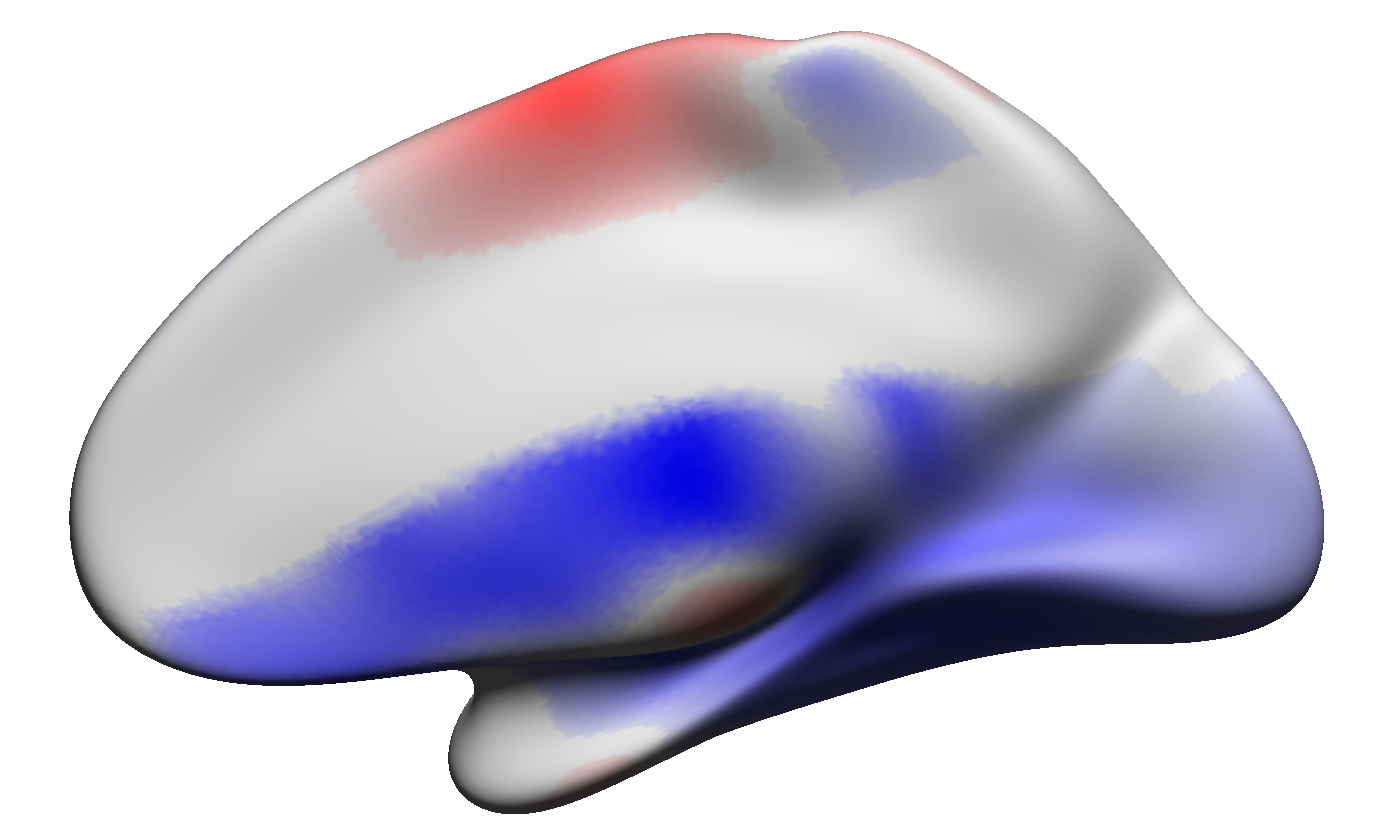}
\includegraphics[width=0.24\textwidth]{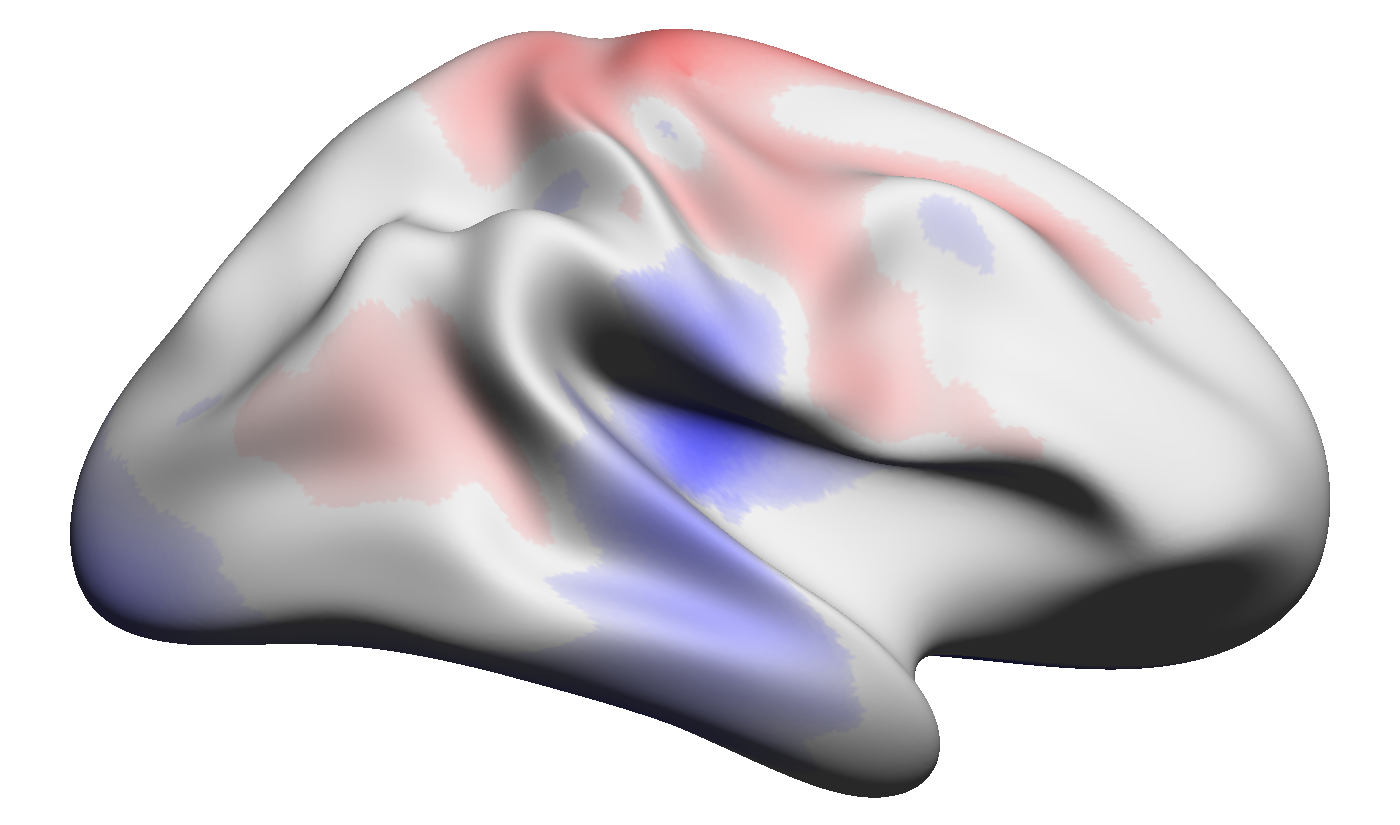}
\includegraphics[width=0.24\textwidth]{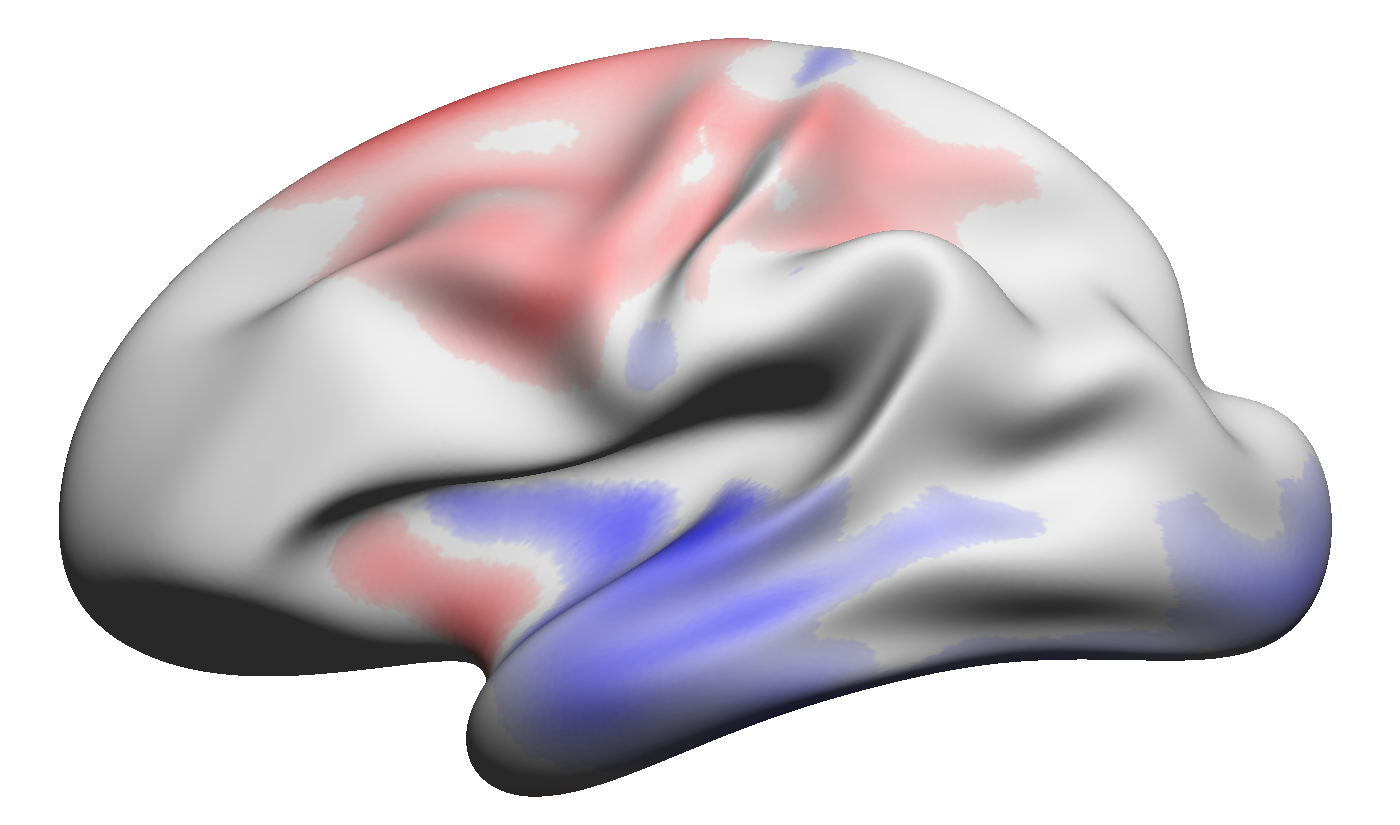}
\includegraphics[width=0.24\textwidth]{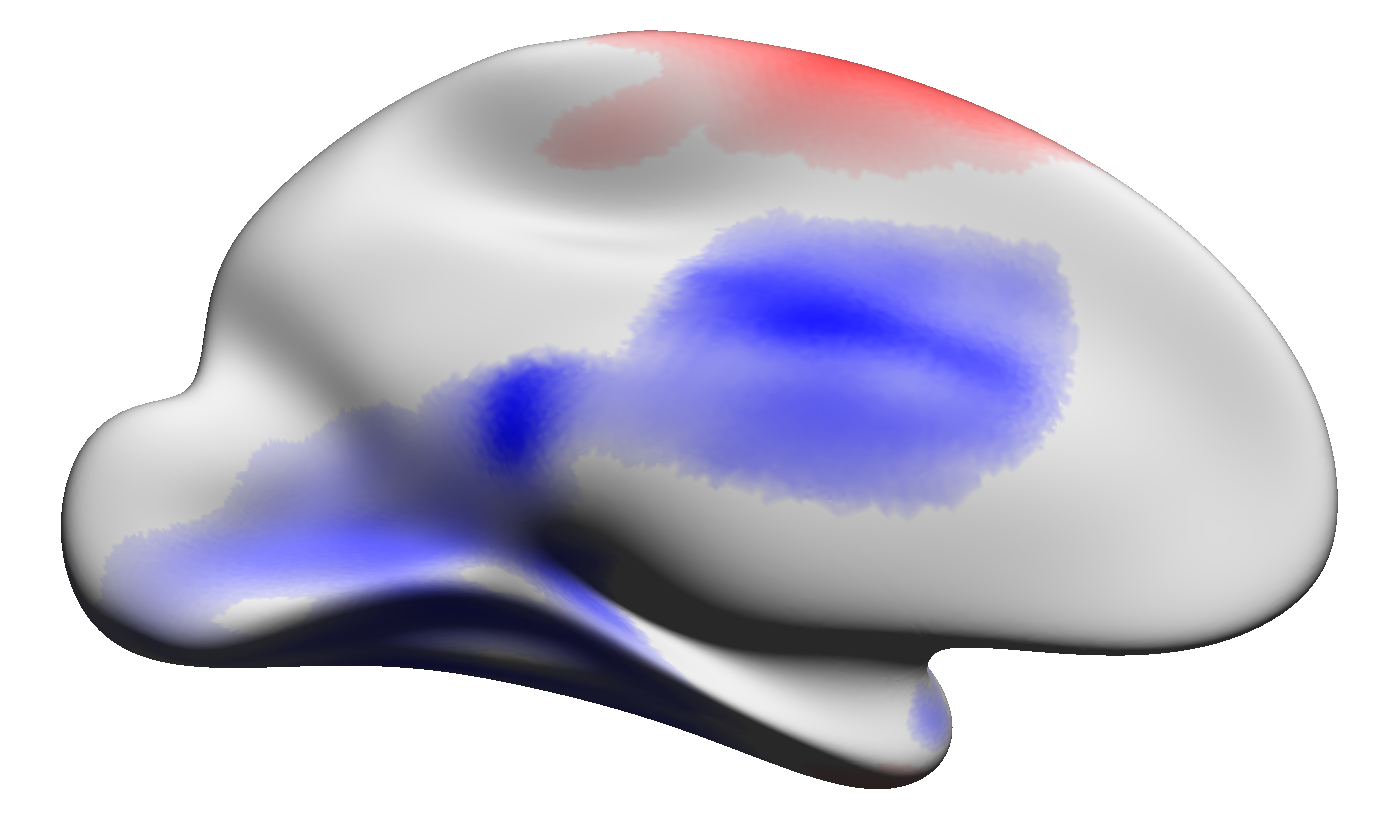}
\includegraphics[width=0.24\textwidth]{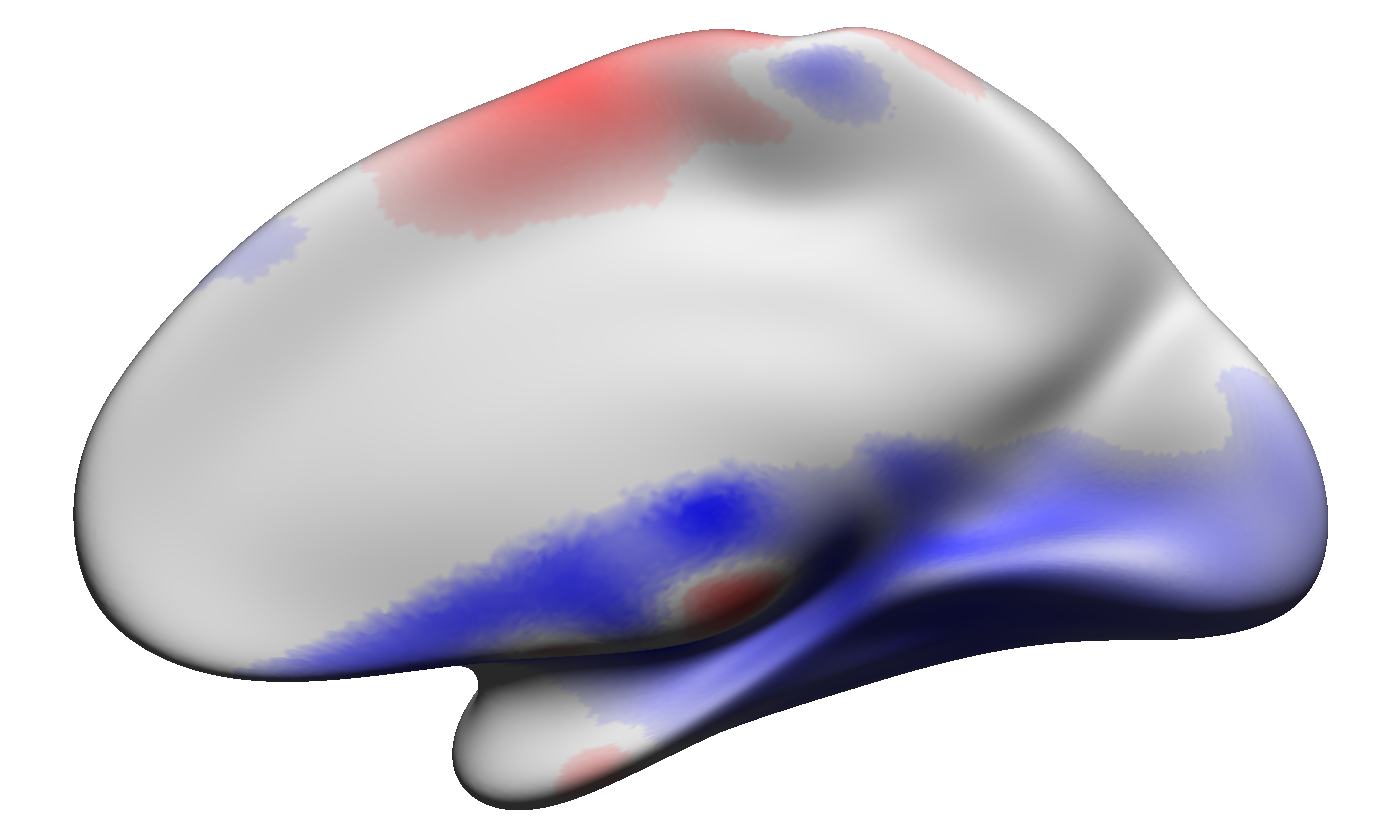}
\includegraphics[width=0.24\textwidth]{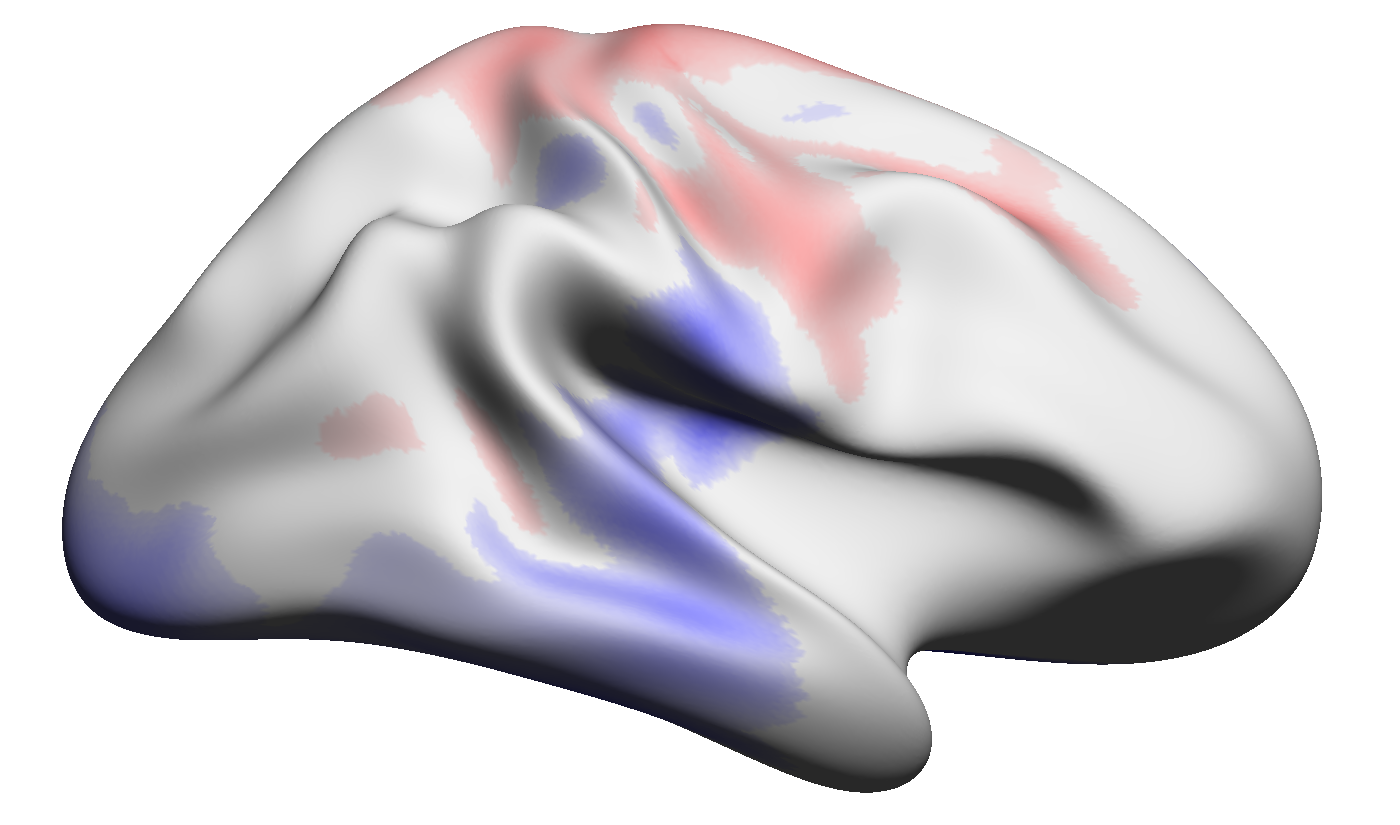}
\includegraphics[width=0.24\textwidth]{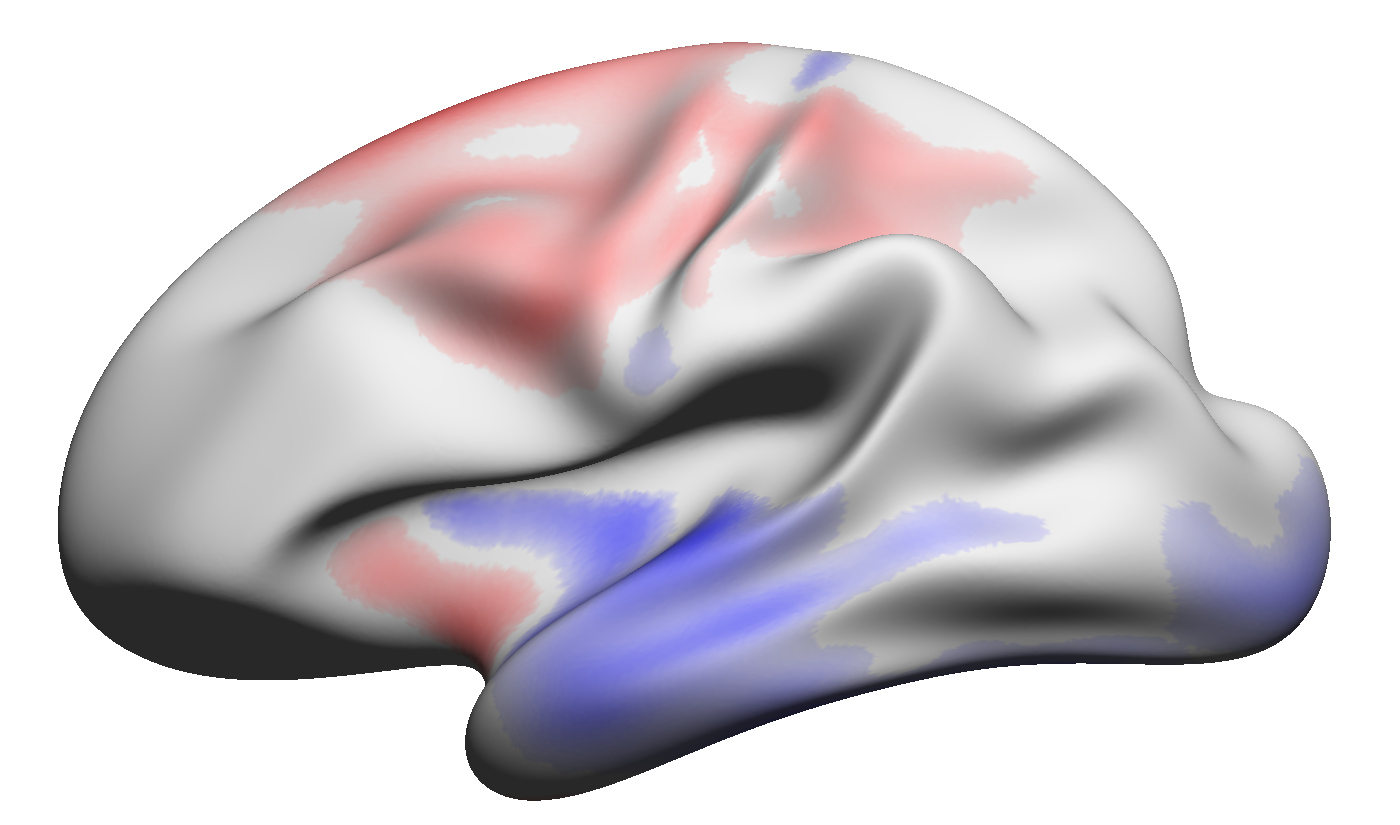}
\includegraphics[width=0.24\textwidth]{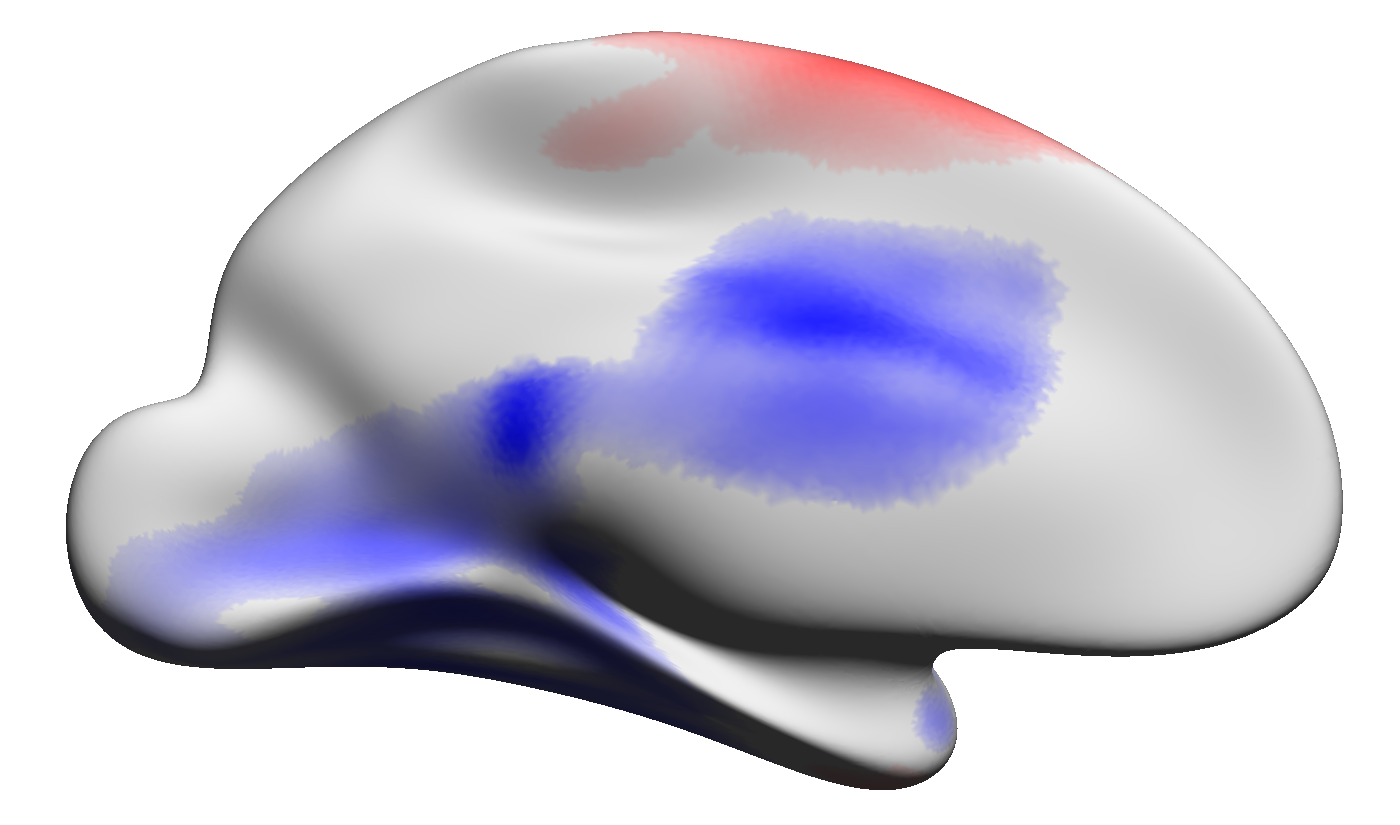}
\includegraphics[width=0.24\textwidth]{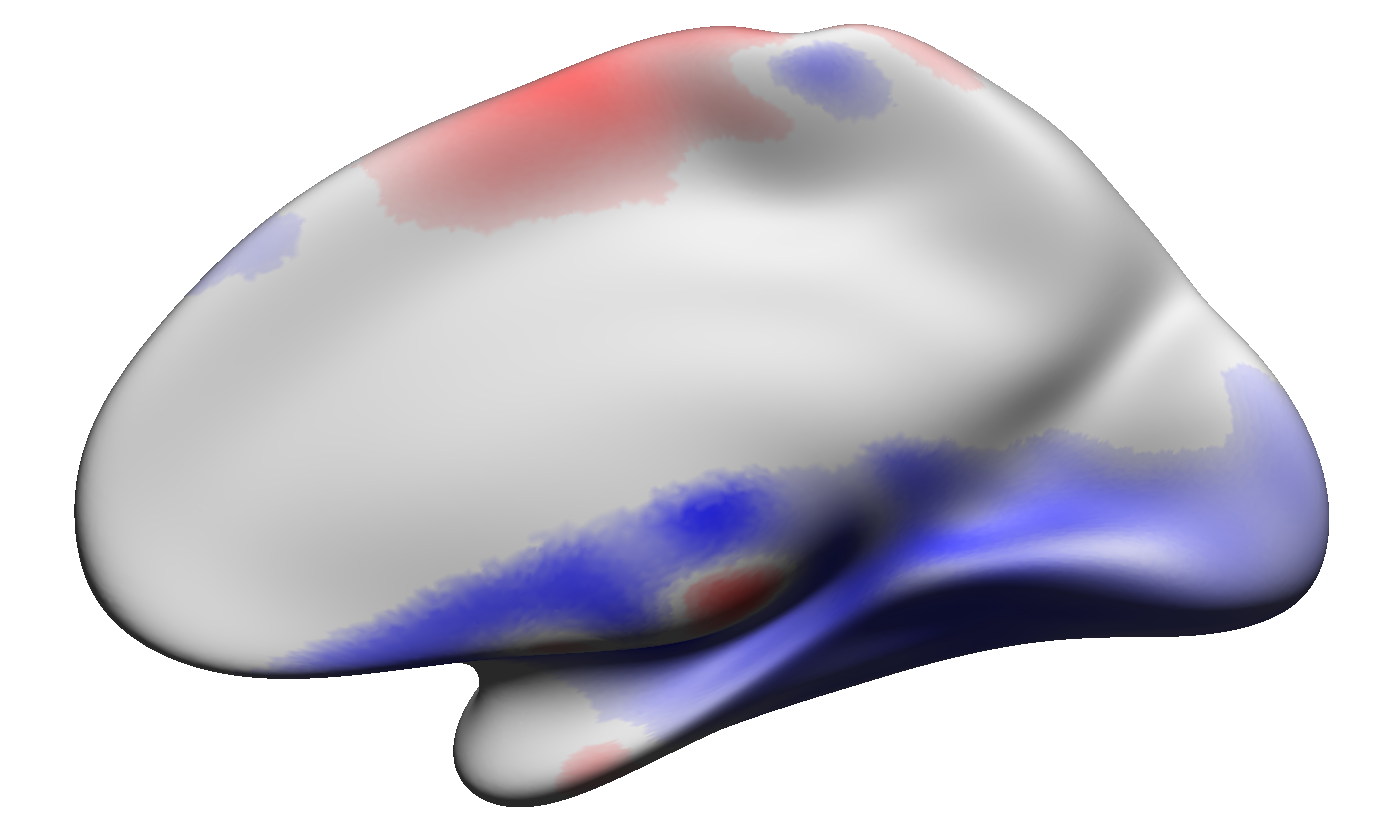}
\includegraphics[width=0.24\textwidth]{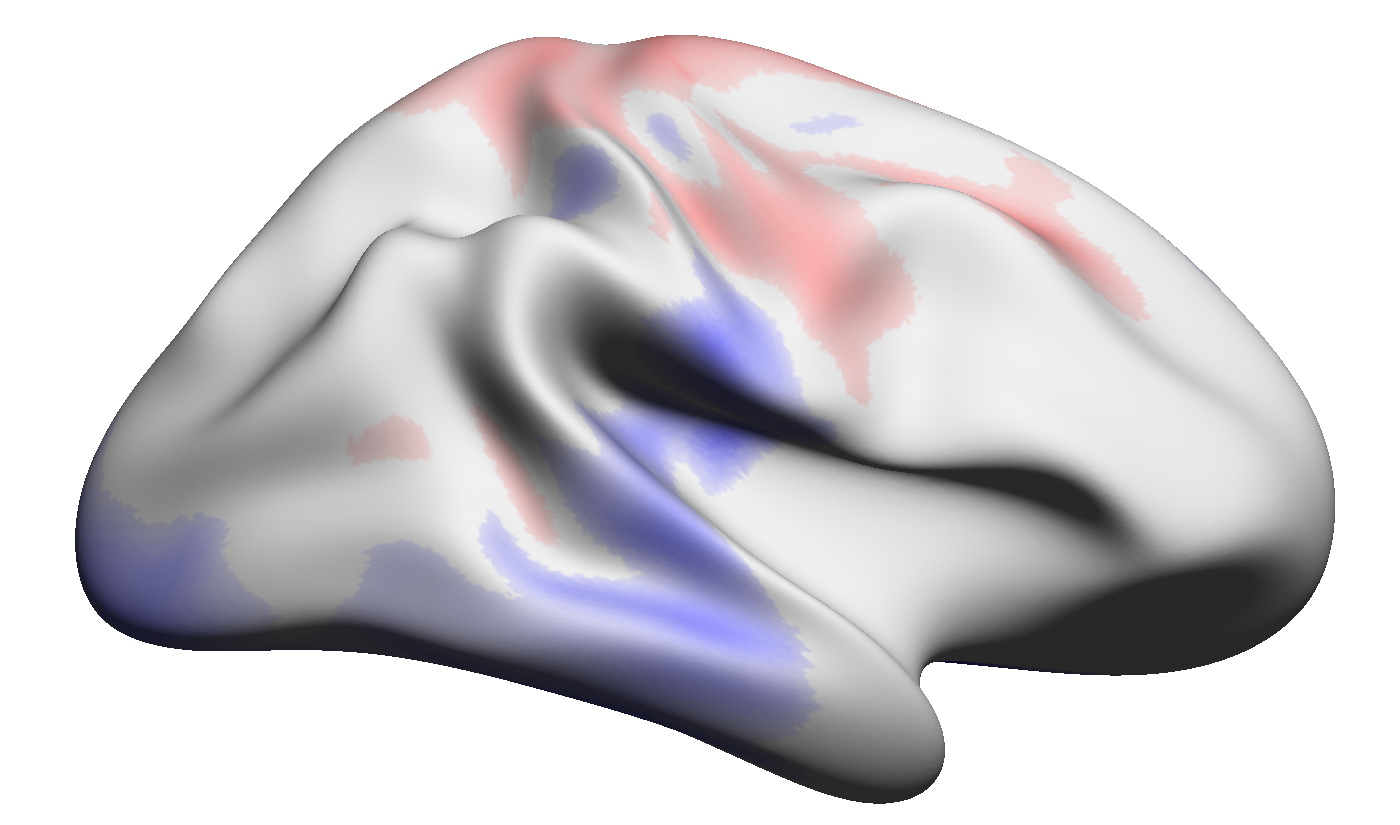}
\includegraphics[width=0.24\textwidth]{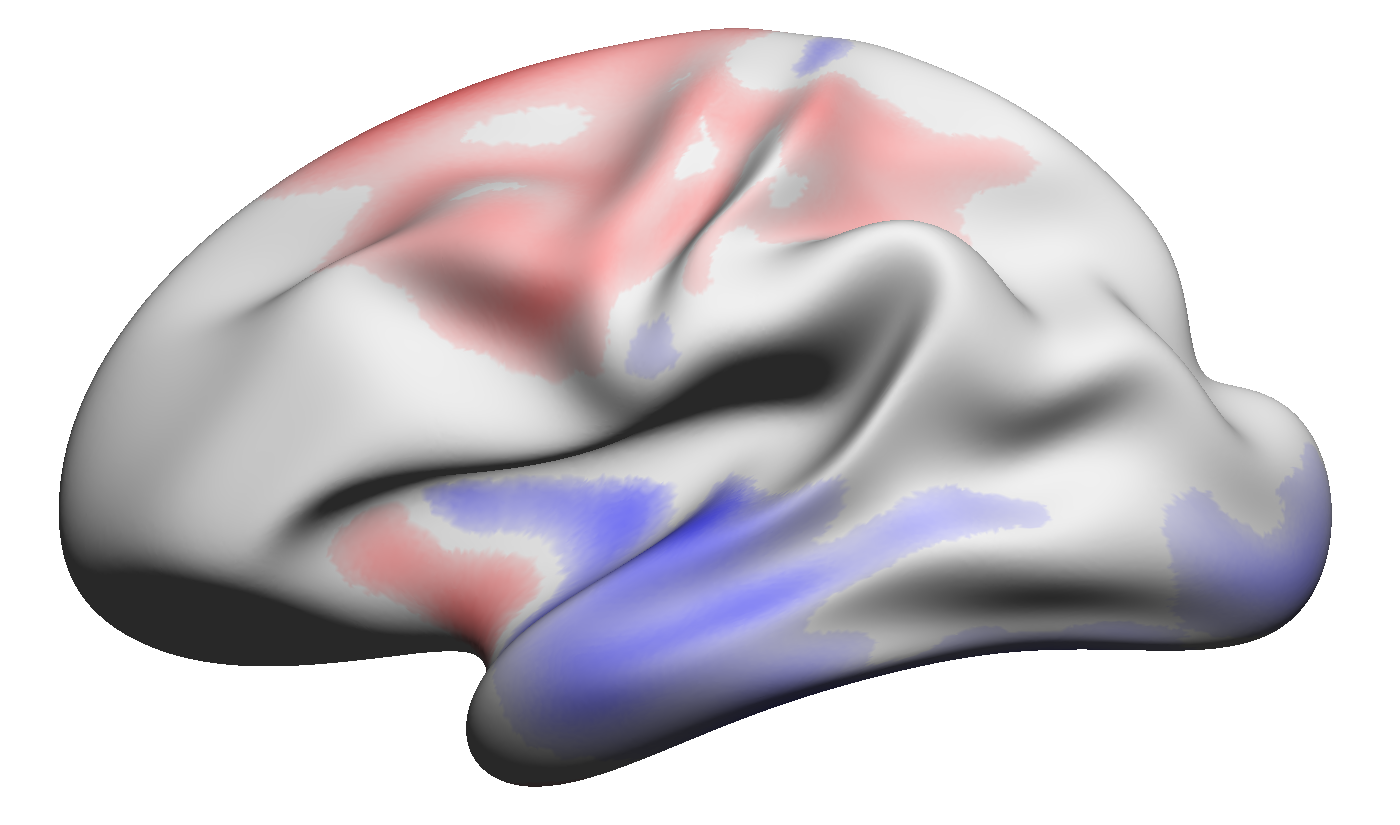}
\includegraphics[width=0.24\textwidth]{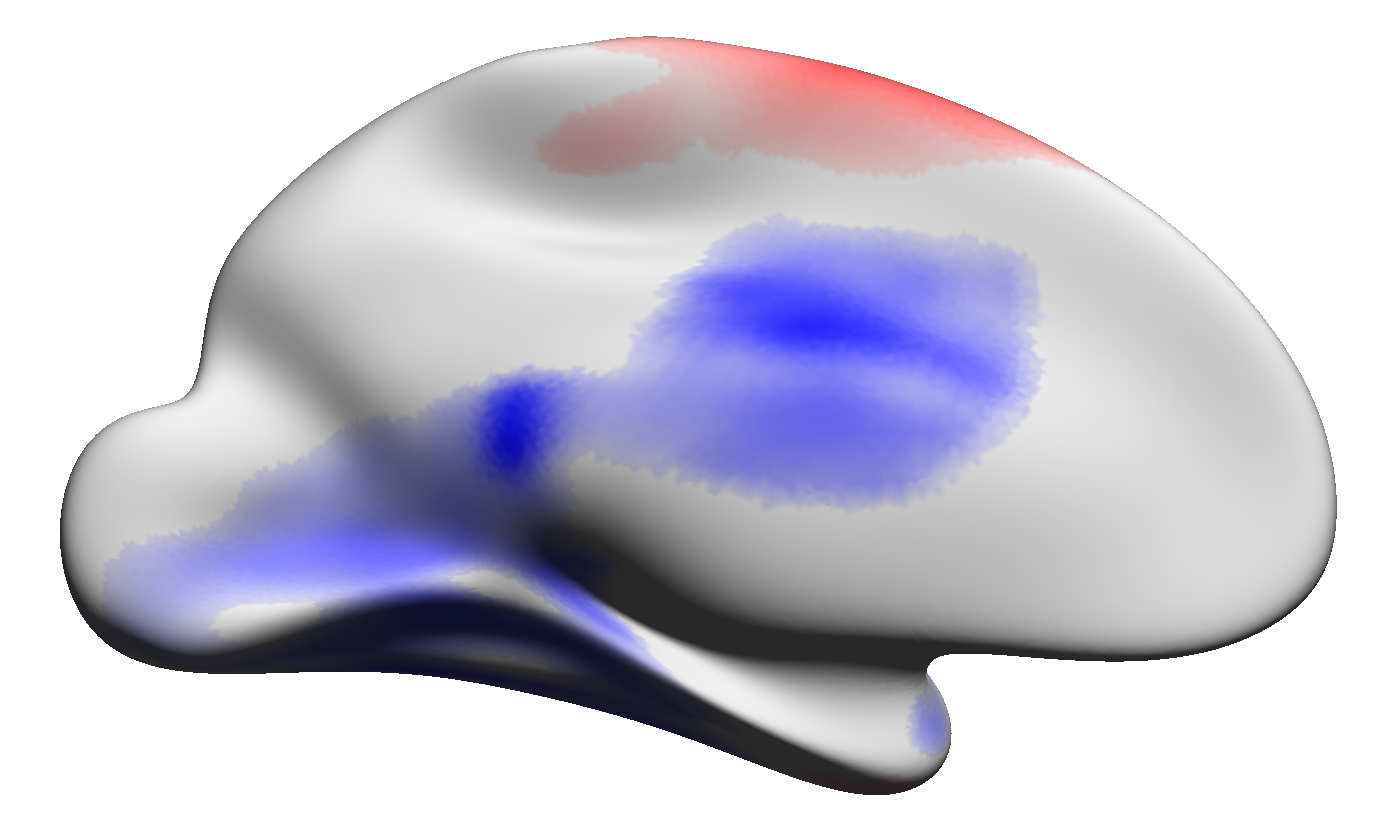}
\includegraphics[width=0.24\textwidth]{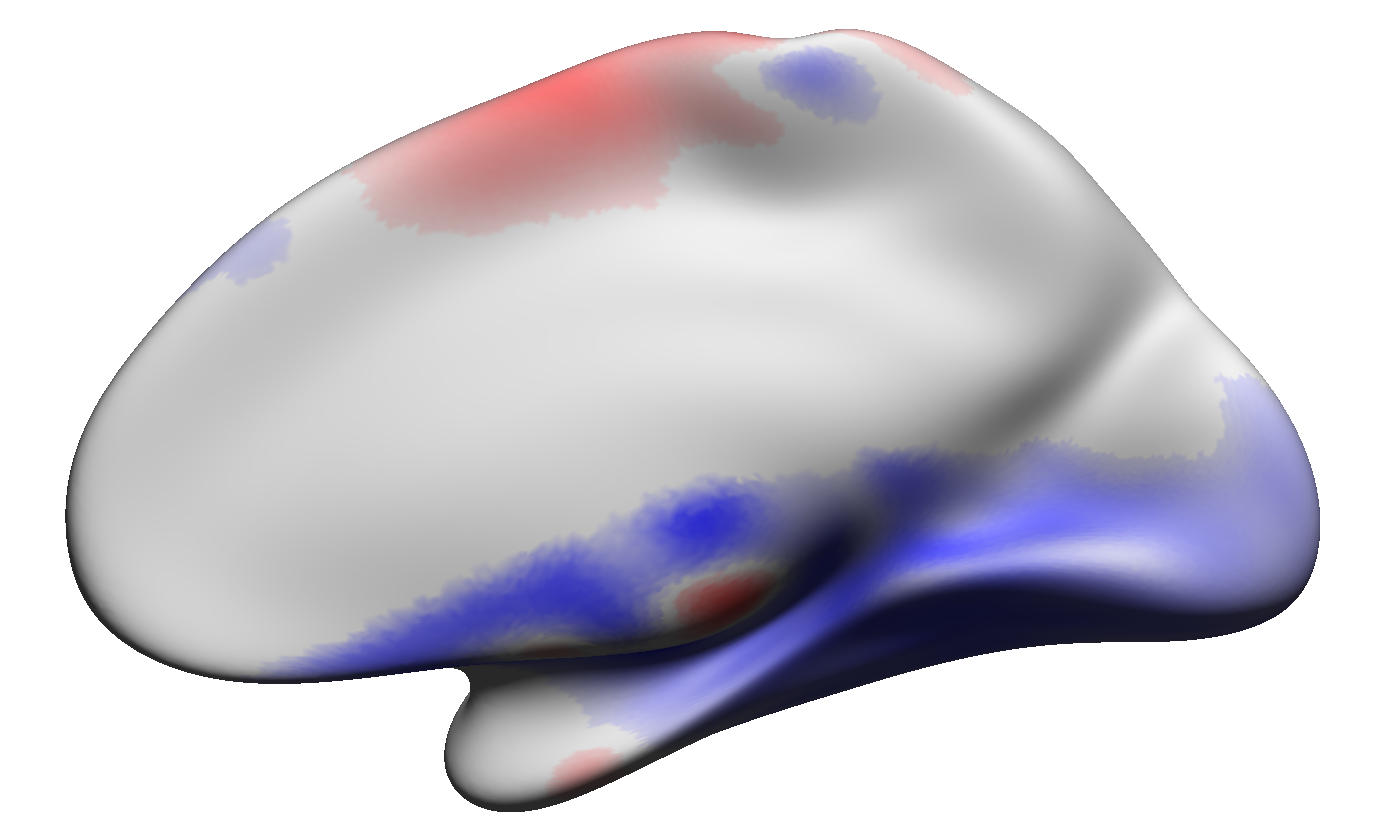}
\includegraphics[width=0.24\textwidth]{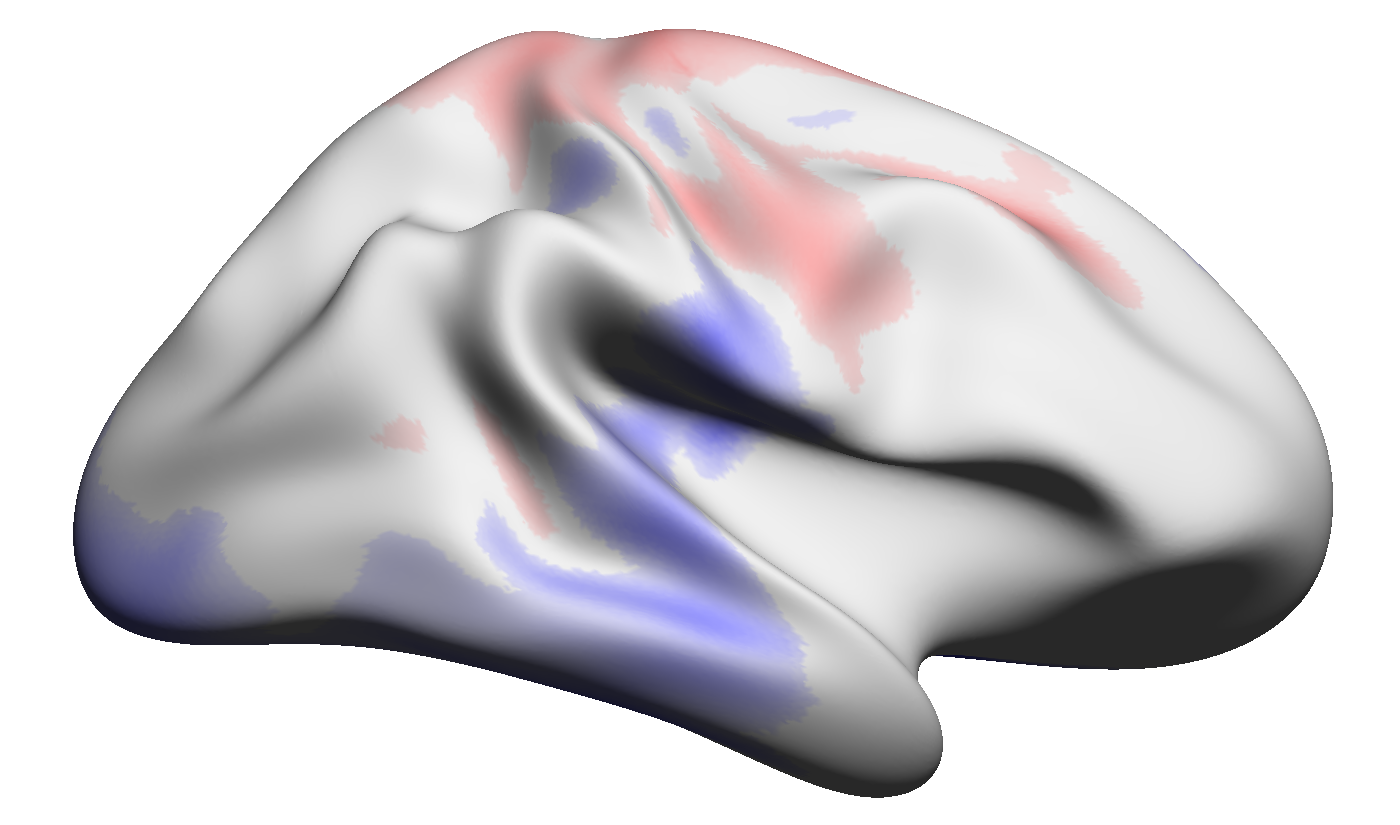}
\caption{In this plot we see p-values for the CSD-DTI test. {\color{blue} Blue }denotes CSD $>$ DTI by as significant margin, while {\color{red} Red} denotes DTI $>$ CSD by a significant margin. Gray areas denote regions that were unable to reject the null hypothesis of ``no difference''. From \textbf{top} to \textbf{bottom} we have $\sigma =\{$0.01,0.005, 0.001, 0.0005, 0.0001$\}$, and from \textbf{left} to \textbf{right} we have the left hemisphere and right hemisphere.}
\label{fig:brains}
\end{figure}

\subsection{Pre-processing}
\label{sec:pre-proc}

Our data are comprised of 731 subjects from the Human Connectome Project S900 release \cite{van2013wu}. We used the minimally preprocessed T1-weighted (T1w) and diffusion weighted (DWI) images rigidly aligned to MNI 152 space. Briefly, the preprocessing of these images included gradient nonlinearity correction (T1w, DWI), motion correction (DWI), eddy current correction (DWI), and linear alignment (T1w, DWI).  We used the HCP Pipeline (version 3.13.1) FreeSurfer protocol to run an optimized version of the recon-all pipeline that computes surface meshes in a higher resolution (0.7mm isotropic) space.  

Tractography was conducted using the DWI in 1.25mm isotropic MNI 152 space. Probabilistic streamline tractography was performed using Dipy's LocalTracking module \cite{garyfallidis2014dipy}.  To model the fiber distribution at each voxel, Dipy's implementation of constrained spherical deconvolution (CSD) \cite{tournier2008resolving} was used with a spherical harmonics order of 8. Tractography streamlines were seeded at 2 random locations in each voxel labeled as likely white matter via the segmentation maps generated by FSL's FAST. Streamline tracking followed directions randomly in proportion to the orientation function at each sample point at 0.5mm steps, starting bidirectionaly from each seed point. Streamlines were only retained if longer than 5mm and both ends terminated in voxels likely corresponding to grey matter according to Dipy's implementation of ACT\cite{smith2012anatomically}. An additional tractography was computed in the same manner, but replacing the CSD fit for a diffusion tensor (DTI) fit.

We then fit 5 separate scales of continuous connectivity models, varying $\sigma$ as to capture differing scales of spatial patterns. We use each $\sigma = \{$0.01,0.005, 0.001, 0.0005, 0.0001$\}$, for both DTI and CSD . We then integrate numerically each resulting intensity function for each subject, forming two marginal connectivity functions for each subject. A paired t-test was then performed at each of the 20484 vertex across subjects (the surface is subsampled for intensity function computation). Multiple comparison correction was applied before inference using the Bonferroni correction for 20484 hypotheses. The continuous assumption makes this correction overly conservative; the test points  are no longer independent due to the smoothness of the signal. Test statistics were evaluated at $\alpha = 0.05$. We also estimate continuous one dimensional densities for samples from the average marginal connectivity function's distribution of values.

\subsection{Degree Sequences}

Much sensation in the past 20 years has focused on complex networks and the distribution of their degree sequences \cite{barabasi2009scale}. We here display estimates for a similar measure for sampled marginal connectivity functions. Multiple studies have provided evidence of small-worldness or scale-free properties in connectomes, both functional and structural \cite{eguiluz2005scale,bullmore2009complex}, sometimes making conflicting claims. While we here make no rigorous claim about the fit of one model or another, we do note the dissimilarity to strictly power law distributions, and their similarity to geometrically constrained communication networks, particularly ad hoc computer networks \cite{hekmat2006connectivity}. (This is qualitative comparison only, and not backed by quantitative observation of ad hoc networks). The plots in Fig \ref{fig:density_ave} show clear second modes, and their log plots are slightly non-linear.

\subsection{DTI-CSD Difference}

As shown in Figure \ref{fig:brains}, there are clear differences recovered by the continuous connectivity model between the CSD and DTI tractography algorithms. This not only illustrates use of a continuous model, but is important for understanding the different trade offs between tractography algorithms. In general DTI is much less flexible than probabilistic CSD, but also fit orders of magnitude faster. Since each connectome is normalized, significant areas are proportionally more explored by a particular algorithm. The authors would also like to reiterate the value and importance of multiple test correction in large hypothesis set situations such as this, due to the large number of sample points tested.

DTI appears to concentrate on motor and somatosensory cortices and their corresponding tracts, while the CSD model appears to find more tracts in the temporal lobe. This is not to say that CSD is missing the motor tracts or that DTI does not have any temporal lobe tracts, but that the relative concentrations are shifted. It is also interesting to note that several small regions break this trend, particularly for DTI concentrations in the inferior temporal and temporal-frotal regions for $\sigma \leq 0.001$. While this work is preliminary, this demonstrates some of the advantages of continuous connectivity models; while region specific models may not be able to resolve such small differences, particularly those surrounded by significant regions of the other density, the continuum model allows very local differences to be detected, as well as large, non-convex regions (such as those in the parietal/frontal regions for DTI, with $\sigma = 0.01$).

This also illustrates one unfortunate issue with the continuum model: Visualization of the full connectivity is difficult (a four dimensional continuum usually embedded in six dimensions and that is \emph{not} a 4-sphere). In future work we plan on addressing this issue and developing methods to visualize portions at a time. However, we also believe this speaks to the possible issues when visualizing the discrete networks.

\section{Conclusion}

In the present work we have described a parcellation free connectivity model. We further used this model to explore degree sequence equivalents for spatial continuum graphs, and to investigate pointwise differences in these functions for two different tractography methods. We believe that parcellation based networks are critical to the exploration of the cortical landscape, but that the development of more general methods such as the one presented here is also vital to expanding the set of testable questions in neuroscience, and improving the answers provided by neuroimaging.

\appendix{}




\bibliographystyle{splncs03}
\bibliography{bacon16.bib}

\end{document}